\documentclass[aps,prb,twocolumn,showpacs,preprintnumbers,amsmath,amssymb,longbibliography,superscriptaddress]{revtex4-2}

\newcommand{\expect}[1]{\langle #1 \rangle}

\newcommand{\be}{\begin{equation}}
\newcommand{\ee}{\end{equation}}
\newcommand{\bea}{\begin{eqnarray}}
\newcommand{\eea}{\end{eqnarray}}

\newcommand{\w}{\omega}
\newcommand{\s}{\sigma}

\usepackage[bottom]{footmisc}
\usepackage{float}
\usepackage[dvipsnames]{xcolor}
\usepackage{ulem}
\usepackage{verbatim} 
\usepackage{amsmath}
\usepackage{graphicx}
\usepackage{dcolumn}
\usepackage{bm}
\usepackage[colorlinks, citecolor={blue!50!black}, urlcolor={blue!50!black}, linkcolor={blue!50!black}]{hyperref}

\definecolor{oucrimsonred}{rgb}{0.6, 0.0, 0.0}

\begin{document}

\title{Underscreened Kondo Compensation in a Superconductor}

\author{Anand Manaparambil}
\email{anaman@amu.edu.pl}
\affiliation{Institute of Spintronics and Quantum Information,
Faculty of Physics, Adam Mickiewicz University,
Uniwersytetu Pozna\'nskiego 2, 61-614 Pozna\'n, Poland}

\author{C\u at\u alin Pa\c scu Moca}
\affiliation{MTA-BME Quantum Dynamics and Correlations Research Group, Eötvös Loránd Research Network,
Budapest University of Technology and Economics, Budafoki út 8., H-1111 Budapest, Hungary}
\affiliation{Department of Physics, University of Oradea, 410087 Oradea, Romania}

\author{Gergely Zar\' and}
\affiliation{MTA-BME Quantum Dynamics and Correlations Research Group, Eötvös Loránd Research Network,
Budapest University of Technology and Economics, Budafoki út 8., H-1111 Budapest, Hungary}

\author{Ireneusz Weymann}
\affiliation{Institute of Spintronics and Quantum Information,
Faculty of Physics, Adam Mickiewicz University,
Uniwersytetu Pozna\'nskiego 2, 61-614 Pozna\'n, Poland}

\date{\today}


\begin{abstract}
A magnetic impurity with a larger $S=1$ spin remains partially screened by the Kondo effect when embedded in a metal.
However, when placed within an $s$-wave superconductor,
the interplay between the superconducting energy gap $\Delta$ and the Kondo temperature $T_K$
induces a quantum phase transition from  
an underscreened doublet Kondo to an unscreened triplet phase,
typically occurring when $\Delta/T_K\approx 1$.
We investigate the Kondo compensation of the impurity spin resulting from
this partial screening across the quantum phase transition,
which together with the spin-spin correlation function
serves as a measure of the Kondo cloud's integrity.
Deep within the unscreened triplet phase, $\Delta/T_K\gg 1$,
the compensation vanishes, signifying complete decoupling of the impurity spin from the environment,
while in the partially screened doublet phase, $\Delta/T_K\ll 1$,
it asymptotically approaches $1/2$, indicating that half of the spin is screened.
Notably, there is a universal jump in the compensation precisely at the phase transition, which we accurately calculate.
The spin-spin correlation function exhibits an oscillatory pattern
with an envelope function decaying as $\sim 1/x$ at short distances.
At larger distances, the superconducting gap induces an exponentially decaying behavior $\sim \exp(-x/\xi_\Delta)$
governed by the superconducting correlation length $\xi_\Delta$,
irrespective of the phase, without any distinctive features across the transition.
Furthermore, the spectral functions of some relevant operators are evaluated and discussed.
In terms of the methods used, a consistent description is provided through the application of multiplicative,
numerical and density matrix renormalization group techniques.
\end{abstract}
\maketitle

\section{Introduction}

\begin{figure}[t]
	\centering
	\includegraphics[width=\columnwidth]{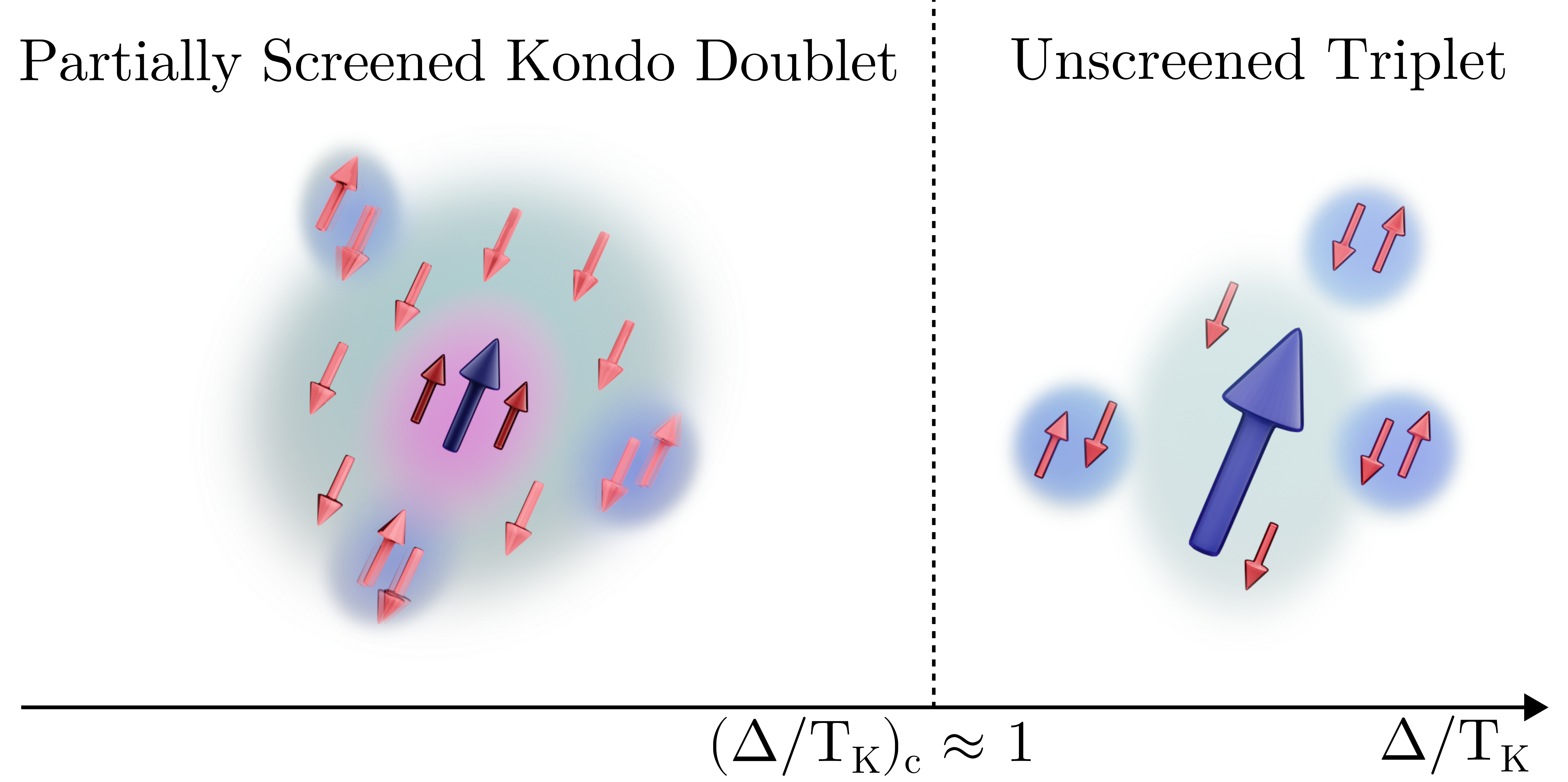}	
	\caption{\label{fig:phase_diagram}
		The illustration of the phase diagram of the underscreened Kondo model.
		For $\Delta/T_K\ll 1$, where $\Delta$ stands for the 
		superconducting energy gap and $T_K$ denotes the Kondo temperature,
		half of the $S=1$ spin is screened by the conduction electrons and the residual spin $S^*=1/2$
		interacts ferromagnetically with the conduction band,
		forming a partially screened doublet phase (underscreened doublet Kondo phase).
		When $\Delta/T_K\gg 1$,
		the impurity spin is decoupled from the underlying superconductor and the ground state is an 
		unscreened triplet.
		}
\end{figure}
	
The Kondo effect is a fascinating and intricate phenomenon that arises in condensed matter physics 
where the magnetic properties of a material, typically a metal, are influenced
by the presence of a magnetic impurity~\cite{DeHaas1936Jun, Kondo1964Jul,Hewson_1993, Goldhaber-Gordon1998Jan, Cronenwett1998Jul, Kouwenhoven2001Jan}. Specifically, when a magnetic impurity is introduced in a non-magnetic host material,
the interaction between the spins of the conduction electrons surrounding the impurity
and the impurity's spin results in a complete or partial screening
of the impurity's magnetic moment~\cite{Nozieres1980Mar, LeHur1997Jul, Sasaki2000Jun, Coleman2003Dec, Posazhennikova2005Jan, Mehta2005Jul, Scott2010Jul}.
As the conduction electrons interact with the impurity,
they form a cloud-like structure around it, known as the Kondo cloud, with its size and properties depending 
on various factors, such as the density of the host electrons
or the local spin-spin exchange coupling~\cite{Chen.1987,Affleck.1996a,Affleck.1996b,Borda.2007,Borda.2009}. 

Recently, by using Fabry-P\'erot interferometry,
the extension of the Kondo cloud has been measured for the first time in the case of a
$S=1/2$ impurity spin~\cite{V.Borzenets2020Mar}.
It indicates that the Kondo cloud can extend over distances of several micrometers,
comparable with the theoretical estimate for the Kondo correlation length $\xi_K\approx v_F/T_K $,
with $v_F$ the Fermi velocity and $T_K$ the Kondo temperature.

When the normal metal turns into a superconductor,
the magnetic impurity disrupts the formation of Cooper pairs
and the screening process competes with the superconducting correlations.
This can give rise to nontrivial effects, such as the emergence of novel electronic states,
the so-called the Yu-Shiba-Rusinov (YSR) bound states \cite{Yu.1965,Shiba.1968,Rusinov.1974},
which appear as subgap excitations in the density of states.
Furthermore, the competition  between the superconducting correlations,
characterized by the superconducting gap $\Delta$, and the Kondo screening, which 
becomes effective below $T_K$, is determining the nature of the ground state.
For an $S=1/2$ impurity spin, when $\Delta/T_K \gg 1$,
the superconducting correlations suppress the screening of the magnetic impurity,
resulting in a  doublet ground state with a decoupled impurity spin from the superconductor.
In the opposite limit, when $\Delta/T_K \ll 1$,
the magnetic impurity locally suppresses the superconducting order parameter,
leading to a singlet many-body ground state,
in which the local spin is screened by the conduction electrons.
Therefore, the system exhibits two distinct quantum phases,
and the corresponding quantum phase transition emerges
when $(\Delta/T_K)_c\approx 1$~\cite{Satori1992Sep, Hatter2015Nov, Kirsanskas2015Dec, Lee2017May}.
So far, this quantum phase transition has been probed experimentally
in various setups \cite{Franke2011May,Deacon2010Mar,Huang2020Dec, Thupakula2022Jun}.

While one may anticipate the existence of the Kondo cloud only in the screened phase,
it has been demonstrated, somewhat counterintuitively,
that the screening cloud persists even when the system is in the doublet phase~\cite{Moca2021Oct}.
A measure, known as the compensation $\kappa$,
\begin{equation}
	\kappa = 1 - \frac{ \expect{S^z}}{S},
	\label{eq:kappa}
\end{equation}
was introduced to assess the resilience of the cloud.
It has been shown that this parameter exhibits a universal jump at the transition~\cite{Moca2021Oct}.

In many cases, impurities carry a larger,  $S>{1\over 2}$, spin.
When introduced in a metal, such an impurity 
experiences the underscreened Kondo effect \cite{LeHur1997Jul,Posazhennikova2005Jan,Mehta2005Jul,Roch2009Nov},
a phenomenon that arises when the number of screening channels $n$
is lower than doubled impurity's spin, $n<2S$ \cite{Nozieres1980Mar}.
When, for example, $n$-channels (assuming $n<2S$) are 
available for screening (e.g. $n$ bands at the Fermi energy),
only part of the impurity spin is screened, leaving behind a residual moment $S^* = S - \frac{n}{2}$. 
The remaining moment $S^*$ couples ferromagnetically to the 
Fermi sea, resulting in a distinctive singular Fermi liquid state~\cite{Mehta2005Jul}.
This state is characterized by a breakdown of the Nozi\` eres' picture of a strong coupling fixed point,
indicating that the essential physics of the underscreened Kondo problem is conceptually different from the regular $S={1\over 2}$ case. 

In the present work, we expand upon our previous study~\cite{Moca2021Oct}
by examining the scenario of a spin $S=1$ magnetic impurity embedded in an $s$-wave superconducting host.
The primary objective of this study is to offer a comprehensive understanding of the fate of the Shiba cloud
in the context of the underscreened Kondo model.
In terms of methods used, we employ a combination of perturbative analytical methods
and non-perturbative numerical techniques based on the renormalization group theory.
More specifically, we construct the phase diagram of the model depicted in Fig.~\ref{fig:phase_diagram}.
Analogous to the conventional $S=1/2$ scenario,
our findings reveal a quantum phase transition from a partially screened (underscreened) doublet Kondo phase
to an unscreened triplet phase, as a function of $\Delta/T_K$.
At the transition point, the compensation $\kappa$ exhibits a universal jump, 
which has been accurately determined by the numerical renormalization group (NRG) method
\cite{Wilson.1975, Bulla.2008}.
Additionally, we compute the spin-spin correlation function in both regimes,
which allows us to determine both the extension as well as the integrity of the Kondo cloud,
together with several spectral functions of the most representative operators.
In our analysis of spatial correlations,
we restrict ourselves to one-dimensional problems,
which enables the application of the density matrix renormalization group (DMRG) method
\cite{White1992Nov, White1993Oct, Schollwock.2005,Schollwock2011}.

\section{Model Hamiltonian and the phase diagram}
\begin{figure}[t]
	\centering
	\includegraphics[width=0.9\columnwidth]{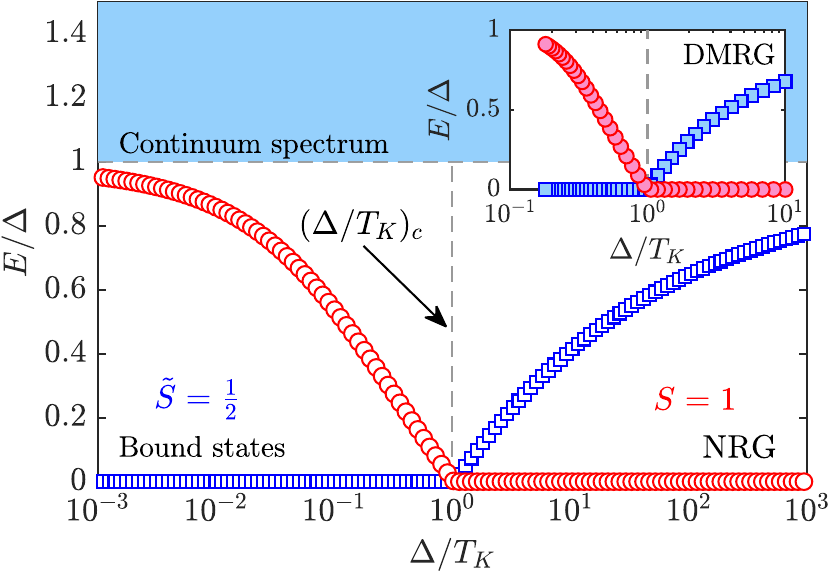}
	\caption{\label{fig:subgap_spectrum}
		The subgap spectrum for the underscreened Kondo model, displaying 
		the evolution of the YSR bound states energy as function of the 
		ratio $\Delta/T_K$. When $\Delta/T_K\ll 1$ the many body ground state is 
		a partially screened doublet. The triplet state resides below the gap and becomes 
		degenerate in energy with the doublet state at the transition. 
		In the opposite limit, when $\Delta/T_K\gg 1$
		the ground state is the unscreened triplet. The vertical line indicated the 
		quantum phase transition that occurs at $(\Delta/T_K)_c\approx 1$. The results in the main 
		panel are obtained using the NRG approach 
		while the inset displays the DMRG results.}
\end{figure}

We consider a spin $S=1$ magnetic impurity embedded in an $s$-wave superconductor.
The Hamiltonian governing the system is given by
\be
H = J\; \vec {\mathbf S} \cdot \vec{\mathbf s} + \sum_{k,\s}\varepsilon_k c^\dagger_{k\s} c_{k\s}
+\sum_{k}\Delta \left( c^\dagger_{k\uparrow} c^\dagger_{-k\downarrow} + {\rm H.c.}\right). 
\label{eq:H_Kondo}
\ee
The first term entails the Kondo exchange interaction between the localized impurity $\vec {\mathbf S}$ and the spin 
$\vec{\mathbf s} = \tfrac{1}{2} \sum_{\s \s'}\sum_k c^\dagger_{k\s} \vec{\s}_{\s\s'} c_{k\s}$
of the conduction electrons, with $\vec{\s}$ denoting the vector of Pauli spin matrices,
and $J$ is the antiferromagnetic exchange coupling.
The second and third terms represent the BCS Hamiltonian characterizing the host superconductor,
with $c^\dagger_{k\sigma}$ ($c_{k\sigma}$) being the creation (annihilation) operators
for an electron of spin $\sigma$, momentum $k$ and energy $\varepsilon_k$.

Constructing the phase diagram of the system
at zero temperature essentially requires understanding how the ground state
and the subgap spectrum evolve as function of $\Delta/T_K$.
To achieve this objective in the most accurate manner,
we use two distinct numerical methods: the numerical renormalization group (NRG)~\cite{Wilson.1975, Bulla.2008}
and the density matrix renormalization group (DMRG)~\cite{White1992Nov, White1993Oct, Schollwock.2005,Schollwock2011}.
In Appendix~\ref{sec:numerics}, further elaboration on both numerical approaches is provided.

The phase diagram of the superconducting underscreened Kondo model is illustrated in Fig.~\ref{fig:phase_diagram},
while the corresponding subgap spectrum is displayed in Fig.~\ref{fig:subgap_spectrum}.
The model exhibits a quantum critical point (QCP) around $(\Delta/T_K)_c\approx 1$,
marking the boundary between two distinct phases.
When $\Delta/T_K\ll 1$, the dominant energy scale is $T_K$, and the local $S=1$ impurity spin is partially
(half) screened by the conduction band electrons.
The residual spin $S^*={1\over 2}$ remains ferromagnetically coupled
with the environment leading to the formation of a doublet ground state $|D\rangle$.
The first excited state takes the form of a YSR triplet state $|T\rangle$
residing within the energy gap. As the ratio $\Delta/T_K$ increases towards $1$,
the triplet state is pulled downward, eventually becoming degenerate with the doublet state precisely at the QCP.
Further increase of the ratio maintains the triplet state as the ground state,
while the doublet state becomes an excited YSR state.
In the triplet phase, the impurity spin decouples from the superconductor,
and the many-body ground state can be envisioned as $|T\rangle \approx |S\rangle \otimes |BCS\rangle$,
where $|S\rangle$ represents the  state of a free $S=1$ spin,
and $|BCS\rangle$ denotes the ground state of the superconductor, hence $|T\rangle$ being triple-degenerate. 

Surprisingly, both the NRG and the DMRG accurately capture
the quantum phase transition and the small discrepancies in the excitation spectrum
are coming from the limitations of each method.
Still, the NRG approach, by design,
proves more suitable for addressing the spin impurity problems,
enabling exploration of the transition across several orders of magnitude
in the $\Delta/T_K$ ratio, $\Delta/T_K\in [10^{-3}, 10^3]$,
while DMRG is able to investigate a narrower window around the transition point
$\Delta/T_K \in [10^{-1}, 10^1]$ due to limitations imposed by the system size.

As an observation, if we would like to compare the phase diagrams
of a fully screened and underscreened Kondo problems, we notice that 
both models exhibit a quantum critical point (QCP) at the same value of $\Delta/T_K\approx 1$.
However, the nature of the two ground states on either side of the QCP is entirely different in the fully
screened~\cite{Bauer.2007,Zitko.2015, Zitko.2016, Moca2021Oct} and underscreened situations.

\section{Compensation and the Kondo cloud}

\begin{figure}[t]
	\centering
	\includegraphics[width=0.9\columnwidth]{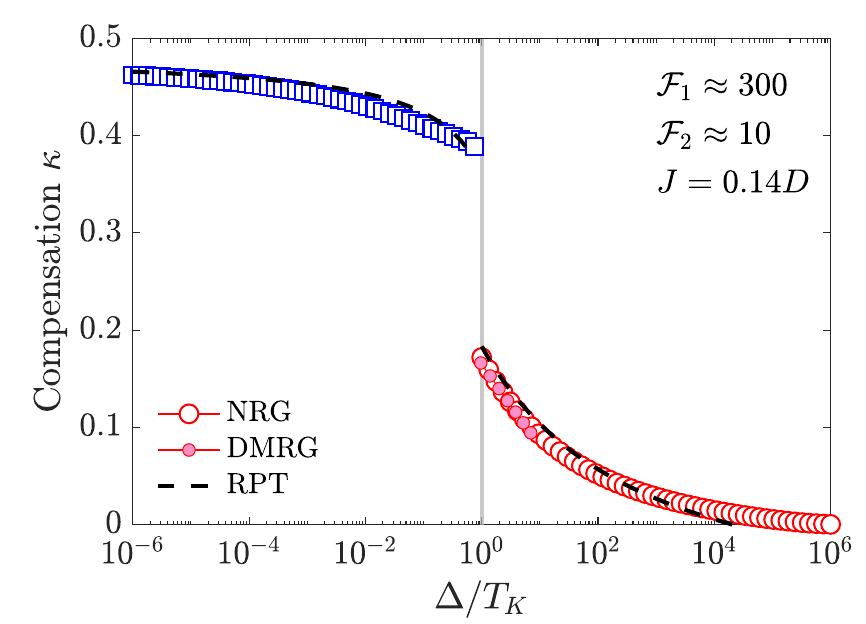}
	\caption{The compensation $\kappa$ as a function
		of $\Delta/T_K$ across the quantum phase transition.
		Empty symbols represent data obtained via the NRG approach,
		while solid symbols depict the DMRG results,
		matching NRG findings in the perturbative regime ($\Delta/T_K>1$).
		The dashed lines denote multiplicative RG (MRG) data.
		Notably, a universal jump in compensation occurs at the QCP,
		approximately $\Delta \kappa \approx 0.22$. }
	\label{fig:compensation}
\end{figure}

The variation of the compensation with respect to $\Delta/T_K$ is depicted in Fig.~\ref{fig:compensation}. 
It displays two limiting values: $\kappa$ tends towards $0$ as $\Delta/T_K$ approaches infinity,
and correspondingly, $\kappa$ approaches $0.5$ as $\Delta/T_K$ tends to $0$.
Remarkably, a universal jump in $\kappa$ occurs at the QCP,
with a value of approximately $\Delta \kappa \approx 0.22$,
derived by using the NRG. The DMRG method also corroborates these findings in the triplet phase,
while overall, the multiplicative RG (MRG) method replicates the otherwise precise NRG results.

We can understand the limiting values of the compensation parameter, $\kappa$,
in the two regimes through straightforward considerations regarding the nature of the ground states.
In the unscreened triplet phase, where $\Delta\gg T_K$,
the impurity spin is decoupled from the superconductor and behaves as a free spin with a $g$-factor, $g=\langle S_z\rangle/S \approx 1$,
indicating no compensation ($\kappa=0$).
As the ratio of $\Delta/T_K$ decreases towards $1$,
the $g$-factor decreases due to the emergence of the Kondo correlations,
resulting in an increase in the compensation. In the opposite extreme limit,
where $\Delta\ll T_K$, $g\approx 0.5$, since half of the impurity spin is screened by the conduction electrons,
implying a compensation value of approximately $\kappa\approx 0.5$ or less.
However, achieving this compensation value numerically poses
a challenge due to the ferromagnetic coupling of the residual spin to the underlying superconductor,
rendering the effective coupling asymptotically approaching zero in a logarithmic fashion.
Consequently, attaining a compensation close to $\kappa \approx 0.5$
typically requires an exceptionally long Wilson chain (or thousands of sites)
to achieve satisfactory convergence in evaluating the $g$-factor.
Details regarding the calculation of the compensation in this limit are discussed in Appendix~\ref{sec:compensation}. 

In the triplet phase, the conventional perturbation theory is effective,
enabling the application of MRG to derive the scaling equations
for the dimensionless coupling and the expression for compensation $\kappa$. Up to 
the third order in perturbation theory, the scaling equation for the effective coupling is
\begin{equation}
	{dj\over dl} = j^2- {1\over 2}j^3+\dots.
	\label{eq:scaling_equation}
\end{equation} 
Remarkably, the RG equations governing the dimensionless coupling $j=J\rho$
remain invariant irrespective of the impurity spin's magnitude.
Here, $\rho= 1/2D$ denotes the constant local density of states,
where $D$ represents half the width of the conduction band,
and $l = \ln (\Lambda_0/\Lambda)$ serves as the scaling variable.
This framework allows us to define the Kondo temperature as
\begin{equation}
T_K\simeq {\cal F}_1 \Lambda_0 \sqrt{j_T}\exp(-1/j_T),\phantom{aaa} j_T>0,
\end{equation}
where $j_T=j(\Lambda\approx\Lambda_0)$ denotes the initial coupling,
and ${\cal F}_2$ acts as an overall parameter ensuring consistency between the Kondo temperature
extracted from NRG data and the perturbative result.

In the doublet phase, on the other hand,
the residual spin $S^*={1\over 2}$ essentially behaves as a free local moment
ferromagnetically coupled to the environment.
In this scenario, the effective coupling is essentially negative.
Solving the same equation~\eqref{eq:scaling_equation} enables the introduction of another characteristic energy scale
\begin{equation}
E_K\simeq {\cal F}_2 \Lambda_0 \sqrt{|j_D|}\exp(-1/j_D),\phantom{aaa} j_D<0,
\end{equation}
which bears a formal resemblance to $T_K$,
albeit it can be linked to an ultraviolet divergence in the scaling equations.
By using the invariance of the free energy under scale transformations,
the $g$-factor and consequently the compensation can be perturbatively determined on both sides of the transition as
\begin{equation}
	\kappa= \begin{cases}
        1- {1\over 2} \exp\left[  \frac {1}{2}(j_D - j(\Delta/E_K)) \right], &  \Delta/T_K < 1 , \\\\
        1- \exp\left[  \frac {3}{2}(j_T - j(\Delta/T_K)) \right], & \Delta/T_K > 1.
        \label{eq:kapp_RG}
    \end{cases}
\end{equation}

Take note of the overall factor within the exponent,
originating from the distinct rescaling of the dimensionless external field.
Additionally, we make the assumption $T_K=E_K$ in our analysis,
a criterion that enables us to naturally merge the two energy scales across the phase transition.

Solving the scaling equations for the coupling in the two regime, the renormalized 
exchange interaction is given by
\begin{equation}
		j (x)\approx \begin{cases} 
			\big[{\ln\left( {\cal F}_2 x \right) + \frac 1 {4  \ln( {{\cal F}_2 x}) }}\big ]^{-1}, &~ x < 1 , \\ \\
			\big[{ \ln ({\cal F}_1 x) - \frac 1 2 \ln \ln ({\cal F}_1 x) + \frac 1 {4  \ln ({\cal F}_1 x )}}\big]^{-1}, & x > 1.
		\label{eq:j}
		\end{cases}
\end{equation}
As can be seen in Fig.~\ref{fig:compensation}, the results for $\kappa$ as obtained from Eq.~\eqref{eq:kapp_RG}
align with the NRG and DMRG data on both sides of the transition, but in order for the perturbation theory to work 
quite large values for the rescaling factors ${\cal F}_{1,2}$ are required.

\begin{figure}[t]
	\centering
	\includegraphics[width=0.9\columnwidth]{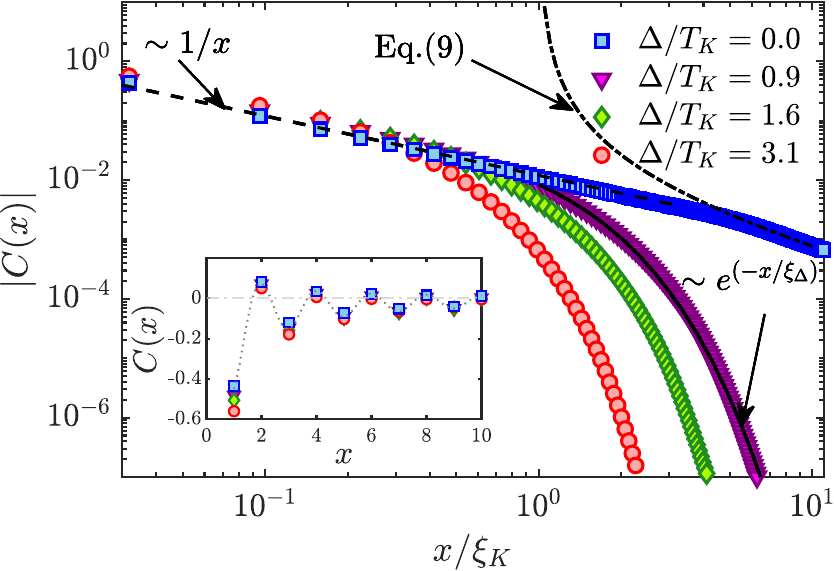}
	\caption{ 
		The envelope of the correlation function $C(x)$
		at the odd sites calculated using DMRG for $J=0.63\, t$ and $L=400$.
		The inset shows the correlation function $C(x)$ for the first $10$ sites.
		The dashed line is the exact analytical result that displays the $2k_F$ oscillation.
	}
	\label{fig:spin_correlation}	
\end{figure}

Thus far, the screening cloud formation has been exclusively explored through compensation measurements.
However, the compensation, while indicative of the Kondo cloud's integrity,
does not offer insight into its real-space structure.
As the itinerant electrons form a cloud around the impurity,
their spins become entangled with the impurity's spin,
leading to a strongly correlated state. The spin-spin correlation function, 
represented as
\begin{equation}
C(\vec r) = \expect {\vec {\bf S} \cdot \vec {\bf s} (\vec r)},
\end{equation}
serves as a tool for characterizing this entanglement, offering further insight into 
the range over which the correlations persist. 
Here, $\vec {\bf S}$ and $\vec {\bf s}(\vec r)$ refer the spin operators
of the impurity and the spin density of conduction band electrons at position $\vec r$, respectively.

For a regular Kondo problem of a spin $S={1\over 2}$ embedded in a normal metal, 
a perturbative analysis~\cite{Ishii1978Aug, Barzykin1998Jan, Sorensen2007Jan}
supported by NRG calculations~\cite{Borda.2007} suggests that the correlation function 
exhibits an oscillatory behavior $\sim \cos(2 \vec {k_F}\vec r) $ with an envelope function that 
displays a power law decay~\cite{Bergmann.2008,Holzner.2009,Dagotto.2010,
Mitchell.2011,Medvedyeva.2013,Lechtenberg.2014,Shchadilova.2014,Debertolis2022Sep,Shim2023Jun}.
For shorter lengthscales  $r\ll \xi_K$, the impurity spin appears as unscreened, and the envelope of the 
correlation function decays as $1/r^d$, where $d$ is the dimensionality of the system.
Conversely, at large distances $r\gg \xi_K$, the impurity is screened and the correlation decays as $1/r^{d+1}$. 
Here $\xi_K$ is the correlation length associated with the Kondo energy scale,
$\xi_K\approx v_F/T_K$, with $v_F$ the Fermi velocity.  

In what follows, we evaluate the spin-spin correlation function using the DMRG in a one-dimensional ($d=1$) chain, for the underscreened Kondo problem. 
More details about the numerical evaluation are outlined in  Appendix~\ref{sec:numerics}.

In the absence of superconducting gap, the decay of the spin-spin correlation function
follows $|C(x)|\sim 1/x$ for distances $x \ll \xi_{K}$,
resembling the behavior of the fully screened Kondo cloud,
as the conduction electrons screen 'half' of the impurity spin.
However, unlike the fully screened Kondo scenario,
the presence of residual moment alters the behavior at larger distances $x \gg \xi_K$,
preserving the $|C(x)|\sim 1/x$ dependence with additional logarithmic corrections.
More precisely, according to Ref.~\cite{Borda.2009}, the envelope of the correlation is given by
\be
|C(x)| \sim \frac{1}{x} \left[ \frac{1}{\ln(x/\xi_K)} - \frac{2}{\ln^2(x/\xi_K)} \right], \phantom{aaa} x \gg \xi_K.
\label{Eq:C_theo}
\ee
With the introduction of the superconducting gap,
excitations at energies below $2\Delta$ are prohibited.
Consequently, the spin-spin correlator experiences suppression beyond the corresponding
correlation length $\xi_\Delta \approx v_F/\Delta$, transitioning from a power-law behavior to an exponential decay: $|C(x)| \sim \exp(-x/\xi_\Delta)$.
This behavior is evident in the DMRG calculations depicted in Fig.~\ref{fig:spin_correlation},
where the numerical findings align well with the analytical predictions.
In the inset of Fig.~\ref{fig:spin_correlation}, the correlation function for the initial
few sites along the chain is displayed, revealing distinct $2k_F$ oscillations,
in agreement with the analytical results~\cite{Borda.2009}.

\section{Spectral functions}

\begin{figure}[t]
	\centering
	\includegraphics[width=0.95\columnwidth]{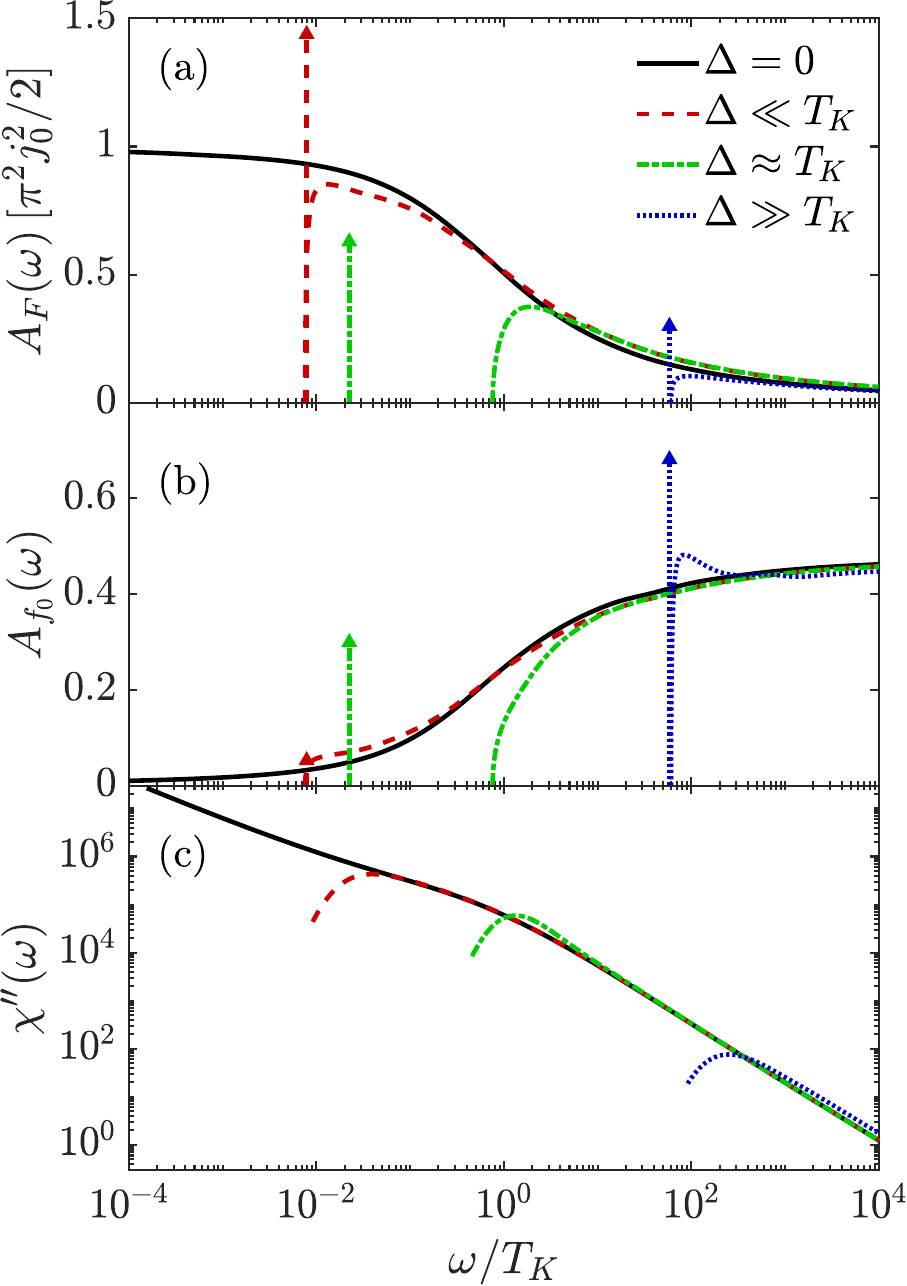}
	\caption{ Spectral functions for several operators computed using the NRG approach for different values of the superconducting energy gap.
		(a) Composite fermion $ F_\sigma = (\vec{\mathbf{S}} \boldsymbol{\sigma} f_0)_\sigma $ (b) the annihilation operator $f_0$ for the first site along the Wilson chain. (c) dynamic magnetic susceptibility $\chi''(\w)$. The calculations are done for $J=0.14 D$ that fixes $T_K = 6.6 \times 10^{-7}$ and different values of the superconducting energy gap $\Delta/T_K = \{0,6 \times 10^{-9}, 6 \times 10^{-7}, 6 \times 10^{-5} \}$. The vertical arrows in panels (a) and (b) correspond to the sub-gap transitions. 
		}
	\label{fig:spectral_functions}
\end{figure}

The presence of subgap states can be detected through transport measurements,
which are closely related to the spectral density of certain specific operators.
For example, the tunneling spectrum of the Shiba state is determined
by the spectral function of the composite fermion operator $ F_\sigma = (\vec{\mathbf{S}} \boldsymbol{\sigma} f_0)_\sigma $.
This operator, characterized by a charge $ Q=1 $ and spin $ S=1/2 $,
shows a gapped spectrum, while the transitions between the in-gap states
are resonances within the gap~\cite{Zitko.2015,Zitko.2017}. This is clearly illustrated in Fig.~\ref{fig:spectral_functions}(a).
Consistent with the spectrum shown in Fig.~\ref{fig:phase_diagram},
the subgap transitions approach the gap energy in both extreme limits of $\Delta/T_K \gg 1$ and $\Delta/T_K \ll 1$.
Additionally, the presence of a resonance near zero energy indicates a degeneracy in the spectrum.
Furthermore, the half-width at half maximum of the spectral function $A_F(\omega) $
in the limit $\Delta = 0$ provides a good estimate of the Kondo temperature $ T_K $. 
In Fig.~\ref{fig:spectral_functions}(b) we display the spectral function
of the annihilation operator $ f_0 $ at the first site along the Wilson chain.
The positions of the in-gap resonances are at the same energies as those corresponding to $ F_\sigma $.

Figure~\ref{fig:spectral_functions}(c) presents the imaginary part
of the spin susceptibility $\chi''(\omega)$.
In the limit of large frequency $\omega/T_K\gg 1$,
the susceptibility $\chi''(\omega)$ shows a decay of the form
\begin{equation}
	\chi''(\omega)\propto {1\over \omega}\,{1\over  \ln^2(\omega/E_K)}.
\end{equation}
Such a singular Fermi liquid behavior has been observed in the ferromagnetically coupled Kondo model~\cite{Kanasz-Nagy2018Apr},
indicating that the unscreened half-spin of the impurity
couples to the conduction band ferromagnetically. For $\Delta =0 $,
the logarithmic dependence persists down to the smallest energies,
while for a finite superconducting gap $\Delta>0$,
all the transition processes with energies below the gap $\Delta$
become suppressed and the susceptibility decays exponentially towards zero for $\omega/\Delta<1$.

\section{Summary}

Our study investigates the behavior of a magnetic impurity with a larger
$ S=1 $ spin embedded in an $ s $-wave superconductor,
concentrating on the interplay between the superconducting correlations,
characterized by the energy gap $\Delta$, and the Kondo correlations,
characterized by the Kondo temperature $ T_K $.
This interplay triggers a quantum phase transition from a partially screened Kondo doublet phase
to an unscreened triplet phase, typically occurring at $\Delta/T_K \approx 1$.
We examined the Kondo compensation of the impurity spin across this quantum phase transition.
In the doublet phase ($\Delta/T_K \ll 1$), the compensation asymptotically approaches $1/2$,
indicating partial screening. On the other hand, in the triplet phase,
around the transition, the compensation is reduced but finite signaling
the existence of remains of the cloud, and vanishes when $\Delta/T_K \gg 1$,
implying a complete decoupling of the impurity spin from the environment.
A universal jump in the compensation at the phase transition point
was observed and accurately calculated.

Additionally, the real space spin-spin correlation function was analyzed,
revealing an oscillatory pattern with an envelope decaying as $\sim 1/x$ at short distances.
At larger distances, the correlation function exhibits an exponentially
decaying behavior $\sim \exp(-x/\xi_\Delta)$, determined by the superconducting correlation length $\xi_\Delta$. 
The spectral functions of several relevant operators have been calculated using the NRG technique,
with some displaying in-gap resonances associated with the transitions between the subgap states.

Our findings were achieved using a combination of multiplicative, numerical,
and density matrix renormalization group techniques,
providing a comprehensive and consistent description of the system.
Thus, our work enhances the understanding of the quantum phase transitions
in the Kondo systems within superconducting environments,
and offers insights into the integrity of the Kondo cloud
and the behavior of spin-spin correlations in these complex systems.

\section*{Acknowledgments}
This work was supported by the Polish National Science
Centre from funds awarded through the decision No.~2021/41/N/ST3/02098, 
by the NKFIH research grants Nos. K134983, K138606, and SNN139581. 
C.P.M acknowledges support from CNCS/CCCDI–UEFISCDI, under projects number PN-IV-P1-PCE-2023-0159. 

\section*{Data availability statement}

The datasets generated and analyzed for this work are publicly available on Zenodo at \href{https://doi.org/10.5281/zenodo.14512777}{https://doi.org/10.5281/zenodo.14512777}.
\appendix

\section{Numerical approaches}\label{sec:numerics}

\begin{figure}[t]
	\centering
	\includegraphics[width=0.9\columnwidth]{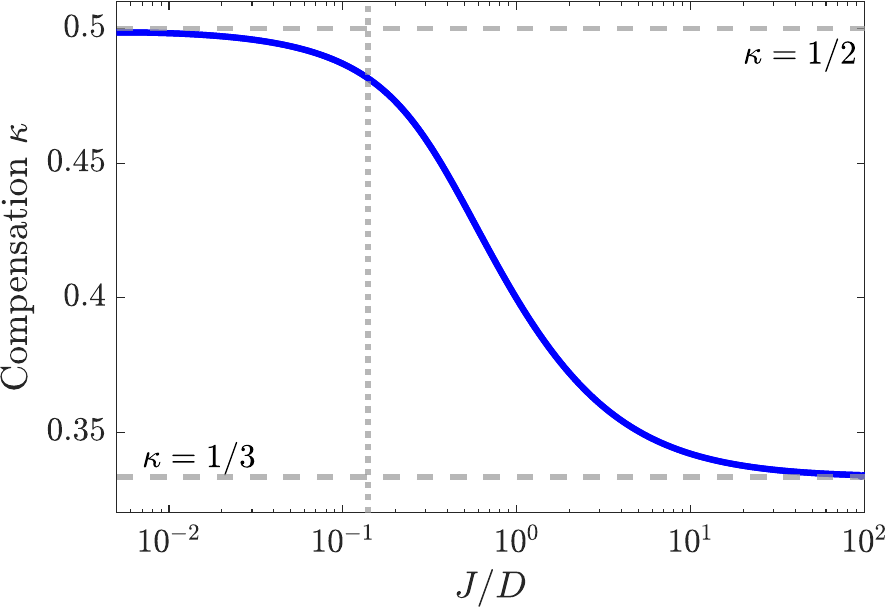}
	\caption{
		The compensation $\kappa$ evaluated via NRG with a zero gap ($\Delta = 0$),
		as a function of $J$.
		The dashed vertical line marks the value of $J=0.14 D$
		employed for computing $\kappa$, as shown in Fig. \ref{fig:compensation}.
		For $\Delta=0$ and $J=0.14D$, $\kappa\approx 0.48$.
}
	\label{fig:kappa_Delta0}
\end{figure}

To construct the phase diagram and calculate the spectral functions and the real-space 
spin-spin correlation function, we make use of two robust numerical methods:
the numerical renormalization group (NRG) 
and the density matrix renormalizationx group (DMRG) approaches. 

Initially pioneered by Kenneth Wilson to investigate the Kondo effect~\cite{Wilson.1975,Wilson.1980, Bulla.2008},
the NRG method involves logarithmic discretization of the system's energy spectrum.
In this approach, the discretized Hamiltonian~\eqref{eq:H_Kondo} undergoes a tridiagonalization procedure
and is subsequently mapped onto a semi-infinite chain, known as the Wilson chain.
In such a mapping, the impurity spin is coupled to the electronic spin density at the first site along the chain,
while the Hamiltonian for the bulk superconductor acquires a one-dimensional tight-binding form
with an on-site pairing correlation.
The hopping along the chain decays with a power-law,
approximately given by $\tau_n \approx\Lambda^{-n/2}$,
where $\Lambda$ is the discretization parameter.
This mapping allows for the systematic truncation of high-energy degrees of freedom,
while preserving the low-energy physics of the original system,
thus allowing for the study of low-energy excitations with high resolution.
When mapped on the Wilson chain, the 
Hamiltonian~\eqref{eq:H_Kondo} becomes 
\begin{eqnarray}
H _{\rm NRG}&=& J\; \vec {\mathbf S} \cdot \vec{\mathbf s} + \sum_{n=0,\s}^{N} \tau_n
\left( f^\dagger_{n,\s} f_{n+1,\s}+ {\rm H.c.} \right) \nonumber\\
&& + \sum_{n=0}^{N} \Delta \left( f^\dagger_{n,\uparrow} f^\dagger_{n,\downarrow} + {\rm H.c.}\right),
\end{eqnarray}
where $f_{n,\s}$ is the annihilation operator of an electron on the chain at site $n$
for spin $\sigma$ and $\vec{\mathbf s} = \tfrac{1}{2}\sum_{\s \s'} f^\dagger_{0,\s} \vec{\s}_{\s\s'} f_{0,\s}$
is the electronic spin operator at the first.
NRG is well suited for evaluating the spectral functions as well as the expectation values of local observables,
allowing us to directly compute the compensation $\kappa$ by evaluating $\langle S_z\rangle $ according to ~\eqref{eq:kappa}.
However, evaluating real space correlations is more involved when employing the NRG~\cite{Borda.2007, Borda.2009}
and for that we utilize the DMRG approach~\cite{White1992Nov, White1993Oct, Schollwock.2005, Schollwock2011}.
In the DMRG approach, we adopt a one-dimensional geometry~\cite{Holzner2009Nov},
in which the impurity spin interacts with the local spin density
at the first site along the chain through an exchange interaction.
Additionally, we assume a uniform hopping along the chain. 
The explicit form of the DMRG Hamiltonian used reads
\begin{eqnarray}
	H _{\rm DMRG}&=& J\; \vec {\mathbf S} \cdot \vec{\mathbf s}(0) - t \sum_{x=1,\s}^{L-1} 
	\left( c^\dagger_{x,\s} c_{x+1,\s}+ {\rm H.c.} \right) \nonumber\\
	&& + \sum_{x=1}^{L} \Delta \left( c^\dagger_{x,\uparrow} c^\dagger_{x,\downarrow} + {\rm H.c.}\right).
\end{eqnarray}
Here, $c_{x,\sigma}$ annihilates an electron at position $x$ for spin $\sigma$,
$t$ denotes the hopping along the chain of length $L$ and 
$\vec{\mathbf s}(x) = \tfrac{1}{2}\sum_{\s \s'} c^\dagger_{x,\s} \vec{\s}_{\s\s'} c_{x,\s}$.
Numerically, we use long chains, $L=200$ or larger,
and in evaluating the ground state of the model and the spin-spin correlations,
we use the matrix product state formalism as implemented in the ITensor library~\cite{itensor, itensor-r0.3}.
We use a bond dimension of $M=512$ or larger,
that guarantees a truncation error in the singular value decomposition (SVD) procedure of $\sim 10^{-8}$ or smaller.

\section{Compensation for $\Delta=0$ and its universality for $\Delta>0$ }\label{sec:compensation}

\begin{figure}[t]
	\centering
	\includegraphics[width=0.9\columnwidth]{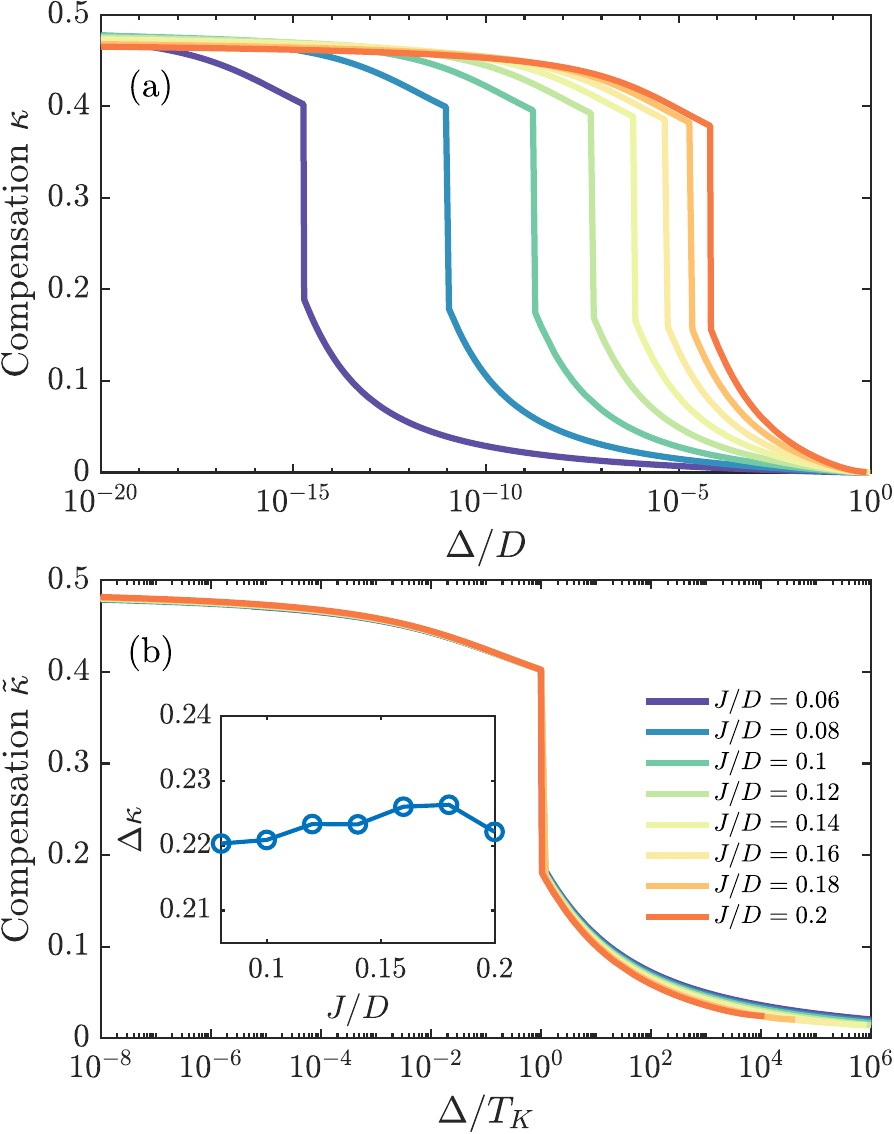}
	\caption{
		(a) The compensation $\kappa$ as a function of $\Delta/D$
		for different values of $J$.
		(b) The  compensation $\tilde \kappa$
		exhibits a universal behavior as a function of $\Delta/T_K$. 
		The inset presents the value of the jump in compensation $\Delta\kappa$ vs $J$.
		The results were obtained by using the NRG approach.
}
	\label{fig:kappa_univ}
\end{figure}

Let us now we examine the value of the compensation $\kappa$ in the limit of $\Delta = 0$.
In the simply minded picture, we expect that in the underscreened Kondo regime, $\kappa = 0.5$,
since half of the impurity spin forms a Fermi liquid with the metallic host,
while the other half remains ferromagnetically coupled to the rest.
However, as shown in Fig.~\ref{fig:compensation},
this is not the case, as $\kappa$ saturates at a value lower than $1/2$ as $\Delta/T_K$ decreases.

This behavior can be understood as follows:
For sufficiently large $J$'s, it is enough to consider just a single site within the Wilson chain.
Then, the expectation value of the impurity's spin can be easily calculated
to be $\langle S_{\rm imp}^z \rangle = 2/3$, resulting in $\kappa = 1/3$.
This is the asymptotic value displayed in Fig.~\ref{fig:kappa_Delta0} in the limit $ J\gg D$.
For lower values of $J$, additional sites of the Wilson chain need
to be considered to accurately capture the Kondo-correlated state,
causing $\langle S_{\rm imp}^z \rangle$ to decrease.
However, $\langle S_{\rm imp}^z \rangle$ can only reach $1/2$ for very small values of $J$,
which correspond to extremely low Kondo temperatures and very long Wilson chains.

This behavior is illustrated in Fig.~\ref{fig:kappa_Delta0}, which shows the NRG results for $\kappa$
as a function of $J$. Note that the values of $J \approx 0.01 D$ correspond to extremely low Kondo temperatures,
requiring very long Wilson chains to capture such energy scales.
To reach this limit we used Wilson chains of length $N \simeq 2000$ with a discretization parameter $\Lambda = 2$.  
The value of $J$ used in Fig.~\ref{fig:compensation} is marked with a dotted
vertical line in Fig.~\ref{fig:kappa_Delta0}, for which $\kappa \approx 0.48$ in the case of $\Delta = 0$.

To demonstrate the universality of the compensation $\kappa$ with respect to the ratio $\Delta/T_K$,
we examine its behavior for various exchange interaction strengths $J$.
In Fig.~\ref{fig:kappa_univ}, the top panel displays $\kappa$ plotted as function of $\Delta/D$.
For each curve, the jump in $\kappa$ occurs at around $\Delta\approx T_K$.
Moreover, the value of $\kappa$ near the transition exhibits a slight dependence on $J$,
consistent with the observations in Fig.~\ref{fig:kappa_Delta0}.
Interestingly, the curves collapse onto a single universal curve when $\kappa$ is plotted as a function of $\Delta/T_K$.
In the inset of the bottom panel, we present the jump in compensation
at the transition $\Delta \kappa$ as a function of $J$.
Remarkably, we find $\Delta \kappa \approx 0.22$ for all values of $J$.

%


\end{document}